\documentclass{IEEEtaes}
\jvol{XX}
\jnum{XX}
\jmonth{XXXXX}
\paper{1234567}
\pubyear{2025}
\doiinfo{TAES.2024.Doi Number}
\usepackage{subcaption,cuted}
\AfterEndEnvironment{strip}{\leavevmode}
\usepackage{amsfonts,amssymb,amsthm,amsmath,mathtools,esint,mwe}
\usepackage{rotating,tabularray,float,cite}
\usepackage{textcomp,verbatim}
\usepackage{graphicx,xcolor}
\usepackage{booktabs}
\usepackage{algorithm,algorithmic}
\begin{document}
\title{Advancing Lunar Communication through Inter-domain Space Networks and Dynamic Orchestration}
\author{Selen Geçgel Çetin}
\member{Member, IEEE}
\affil{Istanbul Technical University, Istanbul, Turkiye\\ 
Polytechnique Montréal, Montréal, Canada}
\author{Barış Dönmez}
\member{Graduate Student Member, IEEE}
\affil{Polytechnique Montréal, Montréal, Canada} 
\author{Güneş Karabulut Kurt}
\member{Senior Member, IEEE}
\affil{Polytechnique Montréal, Montréal, Canada}
\receiveddate{Manuscript received XXXXX 00, 0000; revised XXXXX 00, 0000; accepted XXXXX 00, 0000.\\ 
This work was supported in part by the Tier\,1 Canada Research Chair program and NSERC of Canada Discovery Grant program}
\accepteddate{XXXXX XX XXXX}
\publisheddate{XXXXX XX XXXX}
\corresp{{\itshape (Corresponding author: S.~Gecgel~Cetin)}.}
\authoraddress{S.~Gecgel~Cetin is with Department of Electronics~and~Communication Engineering, Istanbul Technical University, 34469 Istanbul, Turkiye (e-mail:\,\href{mailto: gecgel16@itu.edu.tr}{gecgel16@itu.edu.tr}). S.~Gecgel~Cetin, B.~Donmez and G.~Karabulut~Kurt are with the Poly-Grames Research Center, Department of Electrical Engineering, Polytechnique Montréal, Montréal, QC\,H3T\,1J4, Canada (e-mail:\,\href{mailto:baris.donmez@polymtl.ca}{baris.donmez@polymtl.ca}, \href{mailto:gunes.kurt@polymtl.ca}{gunes.kurt@polymtl.ca}).}
\markboth{Gecgel Cetin \textit{ET AL.}}{Advancing Lunar Communication}
\maketitle
\begin{abstract}
The resurgent era of lunar exploration is defined by a strategic shift from temporary visits to a sustained international and commercial presence, resulting in an unprecedented demand for a robust and continuously available communication infrastructure. The conventional direct-to-Earth communication architecture relies on limited and oversubscribed deep space networks, which are further challenged by the radiative environment and insufficient visibility in certain areas of the cislunar domain. We address these issues by proposing a foundational move toward inter-domain space network cooperation by introducing architectures based on near space networks. They can directly service lunar surface users or, via cislunar relays, by forming a resilient and multi-layered communication backbone. First, we establish a unified link analysis framework incorporating frequently disregarded environmental factors, such as the Moon's variable illumination, to provide a high-fidelity performance evaluation. Second, we assess architectures'~reliability based on the outage risk, essential for quantifying the~operational~robustness of communication links. Finally, to manage the inherent dynamism of architectures, we propose an inter-domain space digital twin: a dynamic decision-making engine that performs real-time analysis to autonomously select the best communication path, ensuring high and stable reliability while simultaneously optimizing power consumption. Overall, our paper provides a holistic architectural and conceptual management framework, emphasizing the necessity of lunar communications to support a~permanent human and economic foothold on the Moon.
\end{abstract}
\section{INTRODUCTION}
T{\scshape he} 21st century marks a new era of lunar exploration by shifting from temporary visits to establishing a permanent and sustainable presence on the Moon~\cite{CH7_1}. A diverse ecosystem of commercial companies and international partners affiliated with long-term ambitions for settlement and resource utilization are all part of this revitalized global endeavor~\cite{Selen1}. This strategic shift from transient missions to sustained presence necessitates continuous operations, a multitude of interacting assets~\cite{CH2_5}, including habitats and rovers~\cite{ref5_t1}, and sophisticated logistical chains~\cite{2ndsub_1}. Therefore, a continuous and reliable communication backbone connecting the Moon to Earth without any interruptions is a vital requirement for these missions, as it has a direct impact on each phase of the mission \cite{CH7_2}. 
\begin{figure*}[!t]
\centering
\includegraphics[width=\linewidth]{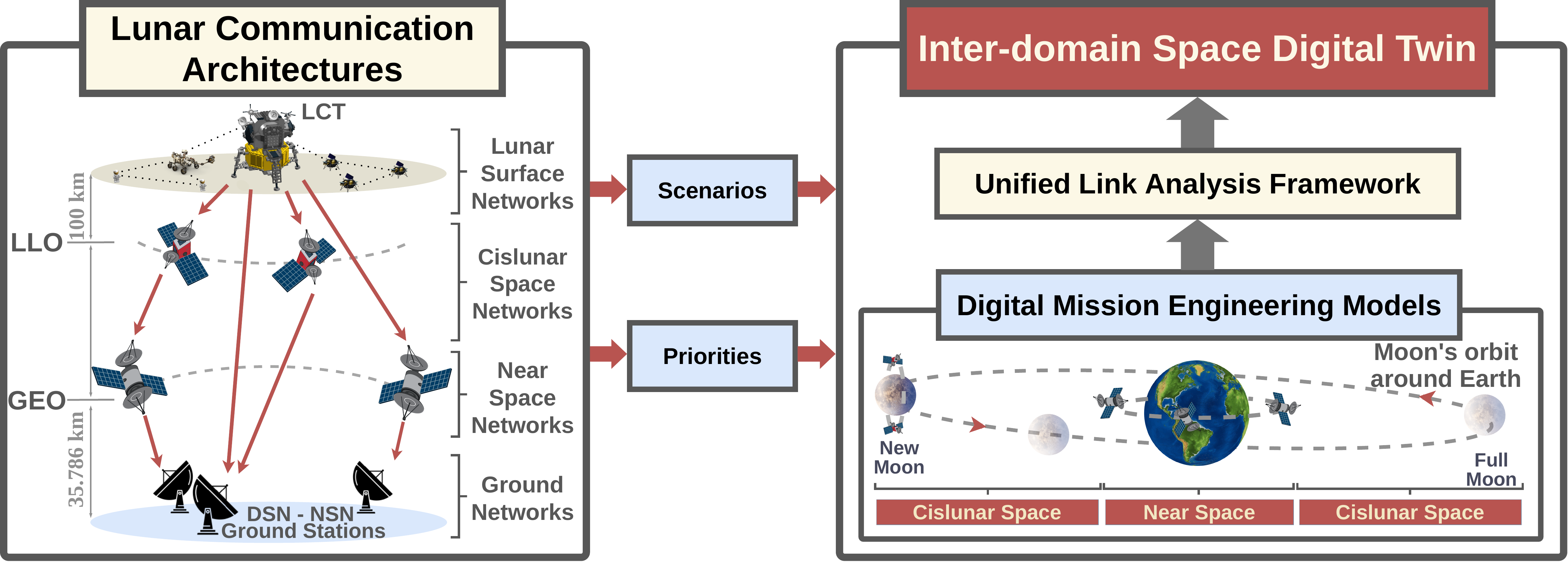}
\caption{System overview for advanced lunar connectivity: from lunar communication architectures to dynamic optimization via an inter-domain space digital twin. The left frame illustrates four lunar communication architectures, from conventional DTE links to multi-hop relaying-based advanced cooperation. On the right frame, the inter-domain space digital twin utilizes possible scenarios and mission priorities (e.g., high reliability or low power consumption) as inputs. The digital twin employs a unified link analysis framework and digital mission engineering models to optimize the entire network for peak performance and reliability by selecting the best communication path.}
\label{Fig1}
\end{figure*}

The conventional way of lunar communication is to establish a \text{direct-to-Earth}~(DTE) link, which is advantageous due to network simplicity and lower latency \cite{ref1_t1}. Despite its simplicity, this solution is accompanied by limitations and challenges that cannot be overlooked, making the conventional communication architecture mismatched to the demands of the new lunar era. The recent attempts by Japan's iSpace highlight these factors that led to the loss of contact with the landers and their eventual crash on the Moon's surface \cite{Japan}, despite both missions being close to achieving a soft landing. Particularly, the visibility of certain areas on the Moon from a ground station significantly varies during a day, resulting in limited line-of-sight~(LoS) availability \cite{CH7_3}. To mitigate this, NASA operates three deep space network (DSN) facilities strategically located about $120^\circ$ apart, but even with this global distribution, any single location on the near side of the Moon has limited and intermittent contact windows~\cite{DSN}. This problem is more critical for the lunar far side, which is permanently shielded from Earth, making DTE communication unfeasible without a relay~\cite{baris1}. 

Establishing a DSN facility is also not economically scalable. To overcome the immense path loss due to the distance, ground stations must be equipped with advanced systems, such as giant antennas \cite{ref4_t1} (such as those with diameters of 70 meters) and high-power amplifiers that transmit signals in the kilowatt range \cite{R1}. This implies that constructing and operating a DSN station can incur costs in the hundreds of millions of dollars, which is why increasing the number of DSN stations for direct communication is not an easily applicable option. Additionally, the DSN is a heavily oversubscribed resource, and the stations are not always used for lunar missions, as they are essentially intended to support deep space missions, creating regulatory constraints \cite{CH7_4} and scheduling conflicts \cite{CH7_5}. Therefore, access to DTE links is challenging for non-agency users such as commercial companies or academic institutions. As these challenges become more apparent with the increasing number of missions, cislunar space networks (CSNs) have been planned by both commercial and national actors to reduce costs, expand access, and ensure reliable communication \cite{Selen2}.

This new paradigm shifts away from centrally managed point-to-point links to a flexible~\cite{CH7_6}, scalable~\cite{CH7_7}, and interoperable~\cite{ref2_t1} \textit{network of networks} such as NASA's emerging framework, LunaNet~\cite{LunaNet}. These solutions are designed to provide the robust and on-demand connectivity required to sustain a permanent presence on the lunar surface by burgeoning multi-layered constellations of relay satellites \cite{CH7_8} and advanced networking protocols~\cite{CH7_9}. CSN architectures in orbits such as low lunar orbit (LLO) \cite{baris2} or near rectilinear halo orbit are proposed to overcome connectivity problems at the Moon~\cite{CH7_10}. However, while CSNs ensure local coverage even for assets on the far side or in shadowed polar regions, a critical architectural question remains: \textit{how do these networks create a robust and efficient transmission back to Earth?} This stimulates inter-domain space networks by facilitating state-of-the-art cooperation among multi-layered CSNs and near space networks (NSNs).

To advance this cooperation between CSNs and NSNs, a more versatile but rational analysis framework is required, given that both cislunar and near space domains\footnote{Near space domain is the operational region extending from the Karman line \cite{CH2_1} to geostationary orbit (GEO) at $35\,786$~km while the cislunar space domain is the volume from GEO to approximately $30\,000$~km beyond the Moon's second Lagrange point \cite{CH2_5}.} have different operational considerations, such as intense traffic in near space \cite{CH2_2} or quite limited sources in cislunar space \cite{CH7_11}. The high congestion of NSNs, especially at lower orbits, and related problems \cite{C1R30} make the concept of a GEO network with two-way communication links more reasonable for inter-domain space networks, as they also provide more stable and wide coverage in both directions. However, the need for multifaceted and generalized link analysis models still exists, while current models are often insufficient because their oversimplified structure fails to capture the unique aspects of each domain and the complex interactions between them \cite{Selen1}. For example, while space links are designed for a dominant LoS, the Moon’s craterous topography introduces significant multipath effects from scattering and reflection, creating a complex channel that must be accurately characterized for reliable link design. Additionally, the harsh radiative environment of deep space, influenced by factors like solar wind and plasma, can degrade signal quality and is often overlooked. A critical oversight in many models is the impact of variable thermal noise \cite{kt1}, where the changing brightness temperature of the lunar surface directly contributes to the system noise temperature of a receiver \cite{kt2} and impacts link reliability by degrading the signal-to-noise ratio (SNR), especially at higher frequency bands like Ka-band \cite{C5R26,ref7_t1,ref8_t1}. This emphasizes the need for a unified link analysis framework that, unlike prior art which often relies on conventional link budgets or models isolated phenomena, holistically integrates these refined environmental effects to provide an accurate and holistic architectural assessment.

The holistic assessments to quantify and mitigate the risk of communication failure under dynamic real-time conditions are mandatory \cite{C4R3}, as a link failure for high-stakes operations is not an acceptable option. This urgency was dramatically highlighted during Astrobotic’s Peregrine mission last year; although a critical propulsion anomaly prevented a lunar landing, its robust communication link was the sole enabler for the ground team to diagnose the problem, salvage valuable data from onboard payloads, and ultimately guide the spacecraft to a safe, controlled re-entry \cite{Mission_fail2}. This situation elevates the importance of rigorous reliability assessment and motivates the outage probability as a critical performance metric to analyze the operational robustness and trustworthiness of critical systems \cite{sennur}. Furthermore, the performance of any given communication architecture—whether DTE, LLO relay, or a hybrid LLO-GEO system—is not always optimal due~to the dynamic interplay of orbital mechanics, network topology, and environmental conditions throughout the missions. Therefore, relying solely on static architectural designs for inter-domain space networks is not entirely satisfactory, which raises a compelling need for dynamic management of these architectures. 

While dynamic management is a mature concept in terrestrial networks, its application to the unique multi-domain (CSN-NSN) cislunar environment, with its stringent reliability demands and complex, high-latency link variations, represents a significant, unaddressed challenge. Such a system must dynamically select the best communication path—switching between lunar communication architectures—to prioritize link reliability while also accounting for domain-level operational limitations like power consumption \cite{CH7_12}. We therefore propose an inter-domain space digital twin. Its decision-making engine uses a high-fidelity virtual model of the network, updated with real-time telemetry, to predict performance, diagnose faults, and autonomously command the network to adapt its routing configuration, ensuring the most reliable and efficient communication path is always active. 

In this paper, we propose to advance lunar communication first through inter-domain space networks by introducing and comparing architectures with practical and dynamical models. We then aim to enhance the utilization efficacy of the architectures with intelligent management. Fig. \ref{Fig1} shows our integrated approach from physical assets to their dynamic orchestration. Our main contributions are listed as follows.
\begin{itemize}
\item We address the severe limitations of conventional lunar communication, which hinder continuous access and mission scalability, by proposing a comprehensive collaboration of inter-domain space networks. We introduce two architectures that integrate a GEO network (in the near space domain) with a LLO network (in the cislunar space domain), enabling more robust and cost-effective communication essential for future lunar endeavors.
\item We present a unified link analysis framework that holistically integrates critical link budget factors by incorporating unique aspects of near and cislunar space. Particularly, our framework includes the transmission dynamics and propagation mechanisms among nodes and a refined characterization of thermal noise sources, such as the variable brightness temperature of the lunar surface (a distinction often overlooked in generalized models). This comprehensive approach provides the essential foundation for precise performance assessment of diverse inter-domain lunar communication architectures.
\item To ensure mission success, we extend our link analysis with a reliability assessment based on outage risk, addressing the paramount need for continuous communication. We formulate the outage probability for introduced architectures to provide vital insights into their operational robustness and trustworthiness under dynamic conditions where communication failures can have severe consequences.
\item To overcome inherent trade-offs in lunar communication architectures and ensure seamless, high-performance communication, we introduce an inter-domain space digital twin. This decision-making engine moves beyond static network designs by offering a dynamic optimization approach that continuously assesses real-time link conditions. It dynamically selects the best scenario by prioritizing reliability while also optimizing power utilization, which is critical for managing the complexity of emerging lunar architectures and enabling autonomous and robust missions.
\end{itemize}

We validate our contributions with presented numerical results that are obtained with the simulations in a well-known digital mission engineering tool, Satellite Tool Kit (STK) \cite{STK_Ansys}. The analysis of in-orbit dynamics across four communication scenarios reveals significant fluctuations in link geometry and visibility, highlighting the inherent unreliability of each architecture. Reliability assessments, based on outage probability, quantify the performance of each architecture, demonstrating that while introduced NSN integrated architectures usually offer advantages, no single option provides continuous, robust connectivity under all dynamic conditions. Conversely, we demonstrate that the proposed inter-domain space digital twin's dynamic orchestration resolves this gap, reducing total daily downtime by over 99.9\% compared to the best-performing lunar architecture alternative. It achieves this by successfully maintaining a stable and highly reliable communication link, even under challenging and variable noise environments, through dynamically selecting the best scenario based on real-time link conditions and expected reliability targets. 
\section{UNIFIED FRAMEWORK FOR LINK ANALYSIS}\label{sec_unified_framework}
To assess the suggested lunar communication architectures depicted in Fig. \ref{Fig1}, we need a detailed analysis system that can represent the complicated and changing environment at different space domains. Conventional link models are frequently insufficient, as their simple structure fails to represent the distinctive features of the cislunar and near space realms. This section, therefore, develops a unified link analysis framework that holistically integrates critical link budget factors to offer accurate analysis. This comprehensive approach provides the essential foundation for the precise performance and reliability assessments conducted in the subsequent sections.
\subsection{Received Signal Power}
The transmitter antenna sends the communication signal to the receiver antenna at the observation planes $(\theta, \phi)$. The received signal power as a function of distance ($d$) between the transmitter and the receiver is given below:
\begin{equation}
P_R (d) = \frac{P_T G_T(\theta_T,\phi_T) G_R(\theta_R,\phi_R)}{ P_{L}(d) L_{T} L_\mathrm{ATM} L_{R}},
\end{equation}
where $P_T$ is the transmit power, $P_{L}$ is the path loss and $G_T(\theta_T,\phi_T)$ and $G_R(\theta_R,\phi_R)$ denote the antenna power gain at the transmitter and the receiver, respectively. The atmospheric loss due to the gaseous attenuation, $L_\mathrm{ATM}$ is included for links with terrestrial networks on Earth. $L_{T}$ and $L_{R}$ are the losses due to transmitter and receiver equipment, respectively. 
\subsubsection{Path Loss}
The locations of a transmitter and receiver are defined as $\left(X_T, Y_T, Z_T\right)$ and $\left(X_R, Y_R, Z_R\right)$ in terms of Earth-Centered cartesian coordinates. The distance between the transmitter and receiver is calculated by considering LoS as follows
\begin{equation}
d=\sqrt{(X_R-X_T)^2 + (Y_R-Y_T)^2 + (Z_R-Z_T)^2}.
\end{equation}
The free space path loss is calculated for each one-way communication link in the end-to-end path as follows
\begin{equation}
P_L (d)= \left ( \frac{4\pi d }{\lambda} \right )^{2},
\end{equation}
where $\lambda$ denotes the wavelength. 
\subsubsection{Antenna Power Gain}
The power gain of an antenna, $G(\theta,\phi)$ is calculated for the corresponding elevation and azimuth angles $(\theta,\phi)$ at the observation planes as
\begin{equation}
G (\theta,\phi) = \eta_o D(\theta,\phi).
\end{equation}
Here, $\eta_o$ is the radiation efficiency\footnote[1]{$\eta_o$ also known as the ohmic efficiency factor or the ohmic-loss factor since it originates from the ohmic-losses. In the case of no antenna losses $\eta_o=1$.} and $D(\theta,\phi)$ is the directivity gain as
\begin{equation}
D(\theta,\phi)= 4\pi \frac{{P}(\theta, \phi)}{\underset{4\pi\;}{\iint} P(\theta, \phi) \sin\theta d\theta d\phi} ,
\end{equation}
where $P(\theta, \phi)$ denotes the antenna power pattern.
\subsubsection{Atmospheric Loss}
The atmospheric loss ($L_\mathrm{ATM}$) is calculated by considering the gaseous attenuation $A_g$ in accordance with the recommendations by the international telecommunication union as below \cite{ITU_P676_13},
\begin{equation}
A_g = \frac{A_o + A_w}{\sin\theta_E}. 
\end{equation}
$A_w$ is total water vapour attenuation and $\theta_E$ is the Earth centered elevation angle. $A_o =\varpi_o h_o$ is total oxygen attenuation where $h_o$ represents the equivalent height attributable to the oxygen and $\varpi_o$ is specific attenuation because of dry air oxygen, nitrogen and Debye attenuation. Note that $L_\mathrm{ATM}$ is not included by setting $L_\mathrm{ATM}=0$ dB in analyses for the links where the receiver is not on the ground.
\subsection{Thermal Noise}
Thermal noise arises from the thermal emissions of electronic devices. It is superimposed on the transmitted signal at the receiver and typically modeled with additive white Gaussian noise (AWGN). To determine SNR, we must characterize the thermal noise sources, which are collectively represented by an operational equivalent noise temperature $T_\mathrm{OP}$. The total noise power is given as follows~\cite{kt1,kt2}
\begin{equation}
N = k T_\mathrm{OP} B,
\end{equation}
where $k$ is the Boltzmann constant $(\approx 1.38 \times 10^{-23} J/K)$ and $B$ is the bandwidth. $T_\mathrm{OP}$ combining all noise contributions is computed as
\begin{equation}
T_\mathrm{OP} = T_\mathrm{SKY} + \Delta T_{A,\mathrm{ext}} + T_{A,\mathrm{int}} + \frac{T_\mathrm{TL}}{\eta_{o}} + \frac{T_{R}}{\eta_{o} \eta_\mathrm{TL}},
\end{equation}
where $T_\mathrm{SKY}$ is the sky noise temperature, $T_R$ is the receiver noise temperature, $ T_\mathrm{TL}$ is the transmission line temperature, $T_{A,\mathrm{int}}$ is internal antenna noise temperature and $\Delta T_{A,\mathrm{ext}}$ is the additional external antenna noise temperature. $\eta_\mathrm{TL}$ shows the thermal efficiency of the transmission line, scaling the noise contributions appropriately for the antenna aperture reference point. Their calculations are detailed as follows. 
\subsubsection{Sky Noise Temperature}
$T_\mathrm{SKY}$ originates from the cosmic microwave background and atmosphere as 
\begin{equation}
T_\mathrm{SKY} = T_\mathrm{ATMP} \left ( 1-\frac{1}{L_\mathrm{ATM}} \right ) + \frac{T_\mathrm{CMB}}{L_\mathrm{ATM}},
\end{equation}
where $T_\mathrm{ATMP}$ is the physical temperature of the atmosphere and $T_\mathrm{CMB}$ is cosmic microwave background noise. 
\subsubsection{Antenna Noise Temperature}
The external antenna noise temperature, $T_{A,\mathrm{ext}}$ arises from the brightness temperature of the subtended body $T_{B}(\theta,\phi)$ and calculated \cite{R3_ITU} as follows 
\begin{equation}
\label{Taexternal}
T_{A,\mathrm{ext}} = \frac{1}{\Omega_{A}} \underset{4\pi \;\,}{\iint} \overline{P}(\theta, \phi) T_{B}(\theta,\phi) d\Omega_A.
\end{equation}
Here, $\Omega_A$ is the antenna beam solid angle \cite{R2_kraus} as
\begin{equation}
\Omega_A = \underset{4\pi \;\,}{\iint} \overline{P}(\theta, \phi) d\Omega_A.
\end{equation}
where $d\Omega_A = \sin(\theta) d\theta d\phi$. $\overline{P}(\theta, \phi)$ denotes the normalized antenna power pattern. In this paper, it is defined for the antennas with the diameter $D_A$ \cite{STK_antenna} as below
\begin{equation}
\overline{P}(\theta, \phi)= 2 \left [\frac{J_1\left ( \frac{\pi D_A}{\lambda} \sin\theta \right )}{\frac{\pi D_A}{\lambda} \sin\theta }\right ]^2. 
\end{equation}

The equation (\ref{Taexternal}) is simplified as $T_{B}(\theta,\phi) = T_{B}$ by assuming that there is no other radiation sources in antenna surrounding and thus the brightness temperature does not change. Then, the increase in $T_{A,\mathrm{ext}}$ as a function of the brightness temperature is calculated for antenna pointed to the bright body as
\begin{equation}
\label{Ta_ex_1}
\Delta T_{A,\mathrm{ext}}=
\begin{cases}
T_{B}, & \Omega_B \gg \Omega_A \\ 
\frac{\Omega_B}{\Omega_A} T_{B}, & \text{ otherwise}.
\end{cases}
\end{equation}
If the antenna is pointed at the lunar surface, the correction factor is added in above equation (\ref{Ta_ex_1}), and it is written as follows:
\begin{equation}
\label{Ta_ex_2}
\Delta T_{A,\mathrm{ext}} = 
\begin{cases}
\frac{1}{2} T_{B}, & \Omega_B \gg \Omega_A\\ 
\frac{\Omega_B}{2\Omega_A} T_{B}, & \text{ otherwise}.
\end{cases}
\end{equation}
The reason for this correction is that the noise sources in radio astronomy exhibit random polarization, which causes the antenna to receive only half of the power \cite{R2_kraus}. The solid angle at the distance $d$ subtended by the bright body (Moon or Earth) is for any $d \geq r_B$ in steradian as follows
\begin{equation}
\Omega_B = 2 \pi \left ( 1 - \sqrt{\frac{d^2 - r_B^2 }{d^2}} \right ),
\end{equation}
where $r_{B}$ denotes the radius of the bright body that is $\approx 1737$ km and $\approx 6371$ km for the Moon and Earth, respectively. $T_{A,\mathrm{int}}$, the internal antenna noise temperature depends on the physical temperature of the antenna $T_\mathrm{AP}$ and is calculated as below
\begin{equation}
T_{A,\mathrm{int}} = T_\mathrm{AP}\left ( \frac{1}{\eta_{o} } - 1 \right ).
\end{equation}
\subsubsection{Transmission Line Temperature}
The transmission line noise temperature depending on the physical temperature of the transmission line ($T_\mathrm{TLP}$) is as follows
\begin{equation}
T_\mathrm{TL} = T_\mathrm{TLP}\left ( \frac{1}{\eta_\mathrm{TL} } - 1 \right ).
\end{equation}
\subsection{Signal and Channel Models}
The calculation of the received signal and noise power requires the consideration of different factors through different architectural scenarios depending on the source and destination: Moon (\text{M}), LLO, GEO, Earth (\text{E}). For the architectural scenario $i \in \mathcal{I}=\{1,2,3,4\} $, the received signal between the source $S \in \left \{ \text{M},\,\text{LLO},\,\text{GEO} \right \}$ and the destination $D \in \left \{\text{LLO},\,\text{GEO},\,\text{E} \right \}$ is defined as follows
\begin{equation}
y_{S,D}^{i} = \sqrt{P_R(d)} h_{S,D}^{i} x + n.
\end{equation} 
Here, $x$ represents the transmitted symbol with $\mathbb{E}[{|x|}^{2}]=1$ and $h_{S,D}^{i}$ is the channel fading coefficient between the source and destination nodes. $n$ is a complex AWGN following $\mathcal{CN}(0,N)$. The instantaneous SNR for each link is computed as
\begin{equation}
\gamma = \frac{P_R(d) \left | h \right |^2}{N}.
\end{equation}

All links originating from the lunar surface are modeled using the Rician fading distribution, as it accurately captures the channel characteristics: a strong, dominant LoS path coexisting with multipath components generated by signal scattering off the Moon's craterous terrain \cite{C5R34}. The Rician $K$-factor provides a physically intuitive parameter to represent the ratio of power between the direct LoS component and these scattered components. While other fading models exist, the Rician distribution offers a more accurate representation than the Rayleigh model, which assumes no LoS path \cite{Multipath1}. It also provides a better balance of accuracy and tractability \cite{RiceK4} for this LoS-dominant scenario \cite{multipath2} compared to more generalized models like the $\lambda-\kappa-\mu$ distribution, a choice well-supported by prior work in the field \cite{Multipath1,multipath2,RiceK4,RiceK1}. The cumulative distribution function of the SNR for a Rician channel is given as:
\begin{equation}
\label{Rician_CDF}
F_\mathrm{Rician}(\gamma)
=1-Q_{1} \left ( \sqrt{2K} , \sqrt{\frac{2 (1+K) \gamma}{\bar{ \gamma}}} \right ),
\end{equation}
where $K$ and $\bar{ \gamma}$ represent the Rician fading parameter and the average received SNR.
\section{RELIABILITY ASSESSMENT FOR LUNAR COMMUNICATION ARCHITECTURES}\label{inter-domain-space-networks}
Building upon the link analysis framework, this section quantifies the reliability of lunar communication architectures in Fig.~\ref{Fig1} for inter-domain space network cooperation. We utilize the outage probability for each network architecture as the reliability performance metric since failures or substantial deterioration in communication quality might compromise mission objectives and even threaten assets or human lives. For any given communication link of $i$-th architecture's scenario, the outage probability is given depending on the minimum SNR threshold ($\gamma_{\mathrm{th}}$) and instantaneous SNR ($\gamma_{S,D}^{i}$) as \cite{alouini}
\begin{equation} 
P_{\mathrm{out}}^i \left(\gamma_{\mathrm{th}}\right) \triangleq \Pr \left( \gamma_{S,D}^{i} \leq \gamma_{\mathrm{th}} \right).
\end{equation}
The outage probability for the first scenario where the information is transmitted directly from \text{Moon-to-Earth} is formulated as
\begin{equation}
\begin{split}
P_{\mathrm{out}}^1 
&= \Pr \left\{ \gamma_{\mathrm{M,E}}^{1} \leq \gamma_{\mathrm{E}}^1 \right\}\\
&= F \left( \gamma_{\mathrm{E}}^1 \right).
\end{split}
\end{equation}

In relaying scenarios, the satellites function as active transponders that utilize the decode-and-forward technique \cite{DF1}. Therefore, the signal is fully regenerated at each hop, making the communication links statistically independent. The total outage probability is thus calculated as the probability that at least one of these independent links fails, causing an outage for the end-to-end chain. For the second scenario, communication occurs from the Moon to a LLO relay, and then from the LLO relay to Earth ground station. As a failure of either the \text{LLO-to-Earth} link or the \text{Moon-to-LLO} link results in an outage for the scenario, the outage probability is expressed as:
\begin{equation}
\begin{split}
\hspace{-0.5mm}P_{\mathrm{out}}^2
&= \Pr \left \{\gamma_{\mathrm{M,LLO}}^{2} \leq \gamma_{\mathrm{LLO}}^2 \right \}
+ \Pr \left \{\gamma_{\mathrm{LLO,E}}^{2} \leq \gamma_{\mathrm{E}}^2 \right \}\\ 
&- \Pr \left \{ \gamma_{\mathrm{M,LLO}}^{2} \leq \gamma_{\mathrm{LLO}}^2, \gamma_{\mathrm{LLO,E}}^{2} \leq \gamma_{\mathrm{E}}^2 \right \}\\ 
&= F\left(\gamma_{\mathrm{LLO}}^2\right) + F \left(\gamma_{\mathrm{E}}^2\right)
- F \left(\gamma_{\mathrm{LLO}}^2 \right) F\left(\gamma_{\mathrm{E}}^2\right).
\end{split}
\end{equation}
\noindent Similarly, for the third scenario, where information is relayed through a GEO satellite, the outage probability is given below depending on the \text{Moon-to-GEO} and the \text{GEO-to-Earth} links.
\begin{equation}
\begin{split}
\hspace{-1.8mm}
P_{\mathrm{out}}^3
&= \Pr \left \{\gamma_{\mathrm{M,GEO}}^{3} \leq \gamma_{\mathrm{GEO}}^3 \right\} 
+ \Pr \left \{\gamma_{\mathrm{GEO,E}}^{3} \leq \gamma_{\mathrm{E}}^3 \right \}\\
&- \Pr \left \{\gamma_{\mathrm{M,GEO}}^{3} \leq \gamma_{\mathrm{GEO}}^3, \gamma_{\mathrm{GEO,E}}^{3} \leq \gamma_{\mathrm{E}}^3 \right \}\\ 
&= F \left(\gamma_{\mathrm{GEO}}^3 \right) + F \left(\gamma_{\mathrm{E}}^3\right)
- F \left(\gamma_{\mathrm{GEO}}^3 \right) F\left(\gamma_{\mathrm{E}}^3\right).
\end{split}
\end{equation}
\noindent The fourth scenario involves a multi-hop path: \text{Moon-to-LLO}, then \text{LLO-to-GEO}, and finally \text{GEO-to-Earth}. The end-to-end outage probability is determined by subtracting the joint success probability of all three links from one as: 
\begin{equation}
\resizebox{\linewidth}{!}{$\displaystyle
\begin{aligned}
P_{\mathrm{out}}^4 
&= \Pr \left \{ \gamma_{\mathrm{M,LLO}}^{4} \leq \gamma_{\mathrm{LLO}}^4 \right \} + \Pr \left \{ \gamma_{\mathrm{GEO,E}}^{4} \leq \gamma_{\mathrm{E}}^4 \right \}\\
&+ \Pr \left \{ \gamma_{\mathrm{LLO,GEO}}^{4} \leq \gamma_{\mathrm{GEO}}^4 \right \}\\
&- \Pr \left \{ \gamma_{\mathrm{M},LLO}^{4} \leq \gamma_{\mathrm{LLO}}^4, \gamma_{\mathrm{GEO,E}}^{4} \leq \gamma_{\mathrm{E}}^4 \right \}\\
&- \Pr \left \{ \gamma_{\mathrm{M,LLO}}^{4} \leq \gamma_{\mathrm{LLO}}^4, \gamma_{\mathrm{LLO,GEO}}^{4} \leq \gamma_{\mathrm{GEO}}^4 \right \}\\ 
&- \Pr \left \{ \gamma_{\mathrm{LLO},GEO}^{4} \leq \gamma_{\mathrm{GEO}}^4, \gamma_{\mathrm{GEO,E}}^{4} \leq \gamma_{\mathrm{E}}^4 \right \}\\ 
&+ \Pr \left \{ \gamma_{\mathrm{M,LLO}}^{4} \leq \gamma_{\mathrm{LLO}}^4, \gamma_{\mathrm{LLO,GEO}}^{4} \leq \gamma_{\mathrm{GEO}}^4 ,\gamma_{\mathrm{GEO,E}}^{4} \leq \gamma_{\mathrm{E}}^4\right \}\\ 
&= 1 - \left(1 - 
F(\gamma_{\mathrm{LLO}}^{4})\right)
\left(1 - F( \gamma_{\mathrm{GEO}}^{4} )\right)
\left(1 - F( \gamma_{\mathrm{E}}^{4} )\right).\\
\end{aligned}$}
\end{equation}
\section{DYNAMIC ORCHESTRATION: INTER-DOMAIN SPACE DIGITAL TWIN}\label{dynamic-dt}
As different CSN architectures provide varying advantages in terms of coverage and link reliability depending on dynamic operational conditions and mission requirements, a mechanism for intelligent and dynamic switching across lunar communication scenarios is critical. Therefore, we propose a digital twin, the dynamic decision-making engine illustrated in the right panel of Fig.~\ref{Fig1}, which models the entire communication system from lunar nodes to ground stations on Earth via LLO and GEO relays.

The primary goal of the digital twin is to optimize communication strategies during missions by balancing the crucial need for reliability with effective resource utilization, particularly in terms of transmit power, which improves overall system performance and enhances adaptability. Here, the inter-domain space digital twin is updated with the status, link conditions, and operational parameters of all relevant nodes each minute. Its comprehensive awareness allows to proactively assess and optimize for continuous communication, striving for stable reliability performance, and efficiency even when faced with potential risks or dynamic channel variations. This capability underpins its function of providing automated and informed decision-making for network operations.
\begin{algorithm}[!t]
\caption{Digital Twin: Power Optimized Inter-domain Space Networks with Reliability Prioritization}
\label{dt_algorithm}
\begin{algorithmic}
\REQUIRE A reserved fallback scenario $i_\text{res} \in \mathcal{I}$;
a set of targeted reliability levels in an order $\mathcal{L}=$~$\left\{\text{High},\cdots,\text{Moderate},\cdots,\text{Low}\right\}$;
maximum acceptable outage probability targets for all $l \in \mathcal{L}$, $P_\mathrm{out}^l$ with $P_\mathrm{out}^\text{High} < \cdots < P_\mathrm{out}^\text{Moderate} < \cdots < P_\mathrm{out}^\text{Low}$.\vspace{0.6mm}
\STATE (1) Initialize link conditions for all scenarios and compute $\gamma$ by using the unified framework of Section \ref{sec_unified_framework}.\vspace{-3mm}
\STATE (2) Compute $P^{i}_\mathrm{out}(\gamma)$ and total transmit power $P^{i}_{T,\mathrm{total}}$ for each scenario.\vspace{0.6mm}
\FOR {each $l \in \mathcal{L}$ (in order from High to Low)}\vspace{0.6mm}
\STATE (3) Determine the set of eligible scenarios:\vspace{-0.5mm}
\begin{equation*}
\bar{\mathcal{I}} = \left\{i \in \mathcal{I} \mid P^{i}_\mathrm{out}(\gamma) \le P_\mathrm{out}^l \right\}.\vspace{-4mm}
\end{equation*}
\IF {$\bar{\mathcal{I}} \neq \emptyset$}\vspace{0.6mm}
\STATE (4) Select scenario ensuring minimum $P^{i}_{T,\mathrm{total}}$ from eligible set:\vspace{-2mm}
\begin{equation*}
i^* = \arg\min_{i \in \bar{\mathcal{I}}~~}
\left\{ P^{i}_{T,\mathrm{total}} \right\}.\vspace{-2.5mm}
\end{equation*}
\STATE (5) Set projected reliability level: $l^* \leftarrow l$.
\STATE \textbf{return} $i^*,~l^*$.
\ENDIF
\ENDFOR
\STATE (6) Defined reliability levels not met:\vspace{-1mm}
\begin{equation*}
i^* \leftarrow i_\text{res}, \quad l^* \leftarrow \text{Fallback Reserved}.\vspace{-3mm}
\end{equation*}\vspace{-2mm}
\STATE \textbf{return} $i^*,~l^*$.
\end{algorithmic}
\end{algorithm}

We present the core logic of the inter-domain digital twin via Algorithm~\ref{dt_algorithm} that hierarchically evaluates available communication scenarios by ensuring that the system always prioritizes achieving the best possible reliability. Initially, the algorithm assesses link conditions for all scenarios using the unified framework (Section~\ref{sec_unified_framework}) to compute their respective outage probabilities and required total transmit power. It then iterates through predefined reliability levels from highest to lowest. For each level, it identifies scenarios meeting the outage target and, crucially, selects the one that minimizes total transmit power. If no scenario meets even the low reliability requirement, a predefined fallback scenario is activated as baseline connectivity. In this way, it ensures reliable communication, optimizes power consumption, and automates complex operational decisions by dynamically selecting the most suitable scenario based on current conditions and mission priorities. We will further demonstrate the practical efficacy of this digital twin in Section~\ref{numeric_res}.
\begin{figure*}[!t]
\centering
\begin{subfigure}{0.245\linewidth}
\centering
\includegraphics[width=\linewidth]{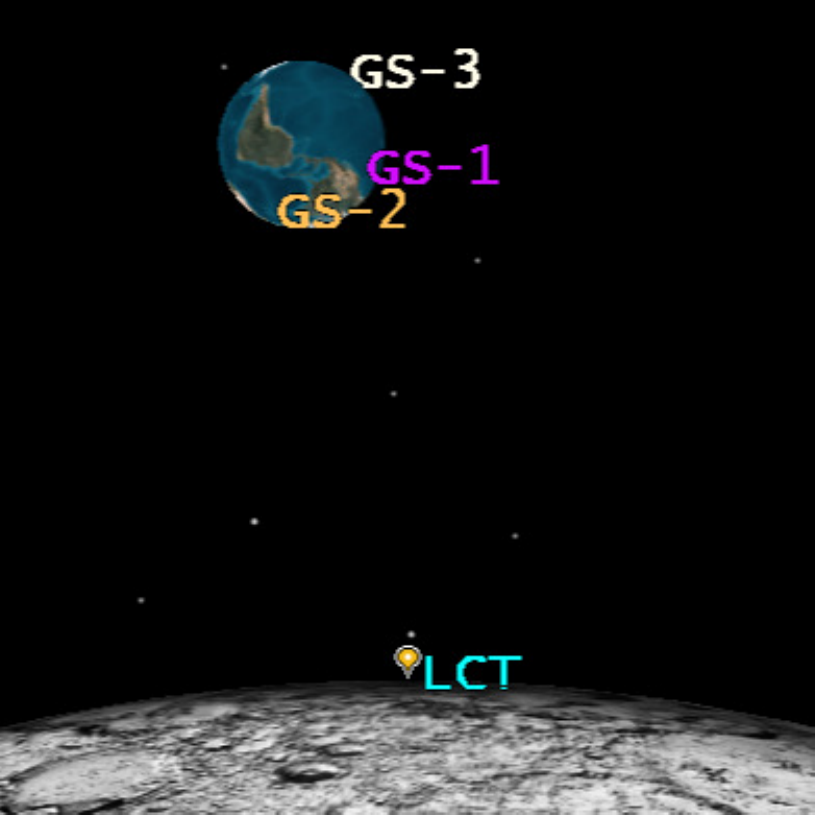}
\caption{Scenario~1}
\label{subfig_a}
\end{subfigure}\hspace*{\fill}
\begin{subfigure}{0.245\linewidth}
\centering
\includegraphics[width=\linewidth]{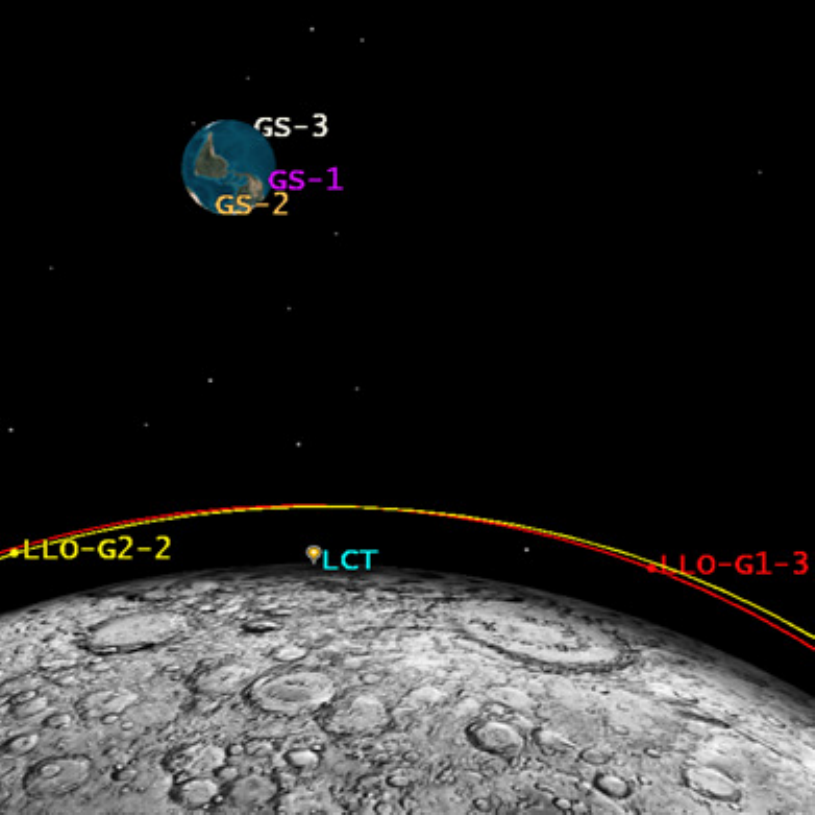}
\caption{Scenario~2}
\label{subfig_b}
\end{subfigure}\hspace*{\fill}
\begin{subfigure}{0.245\linewidth}
\centering
\includegraphics[width=\linewidth]{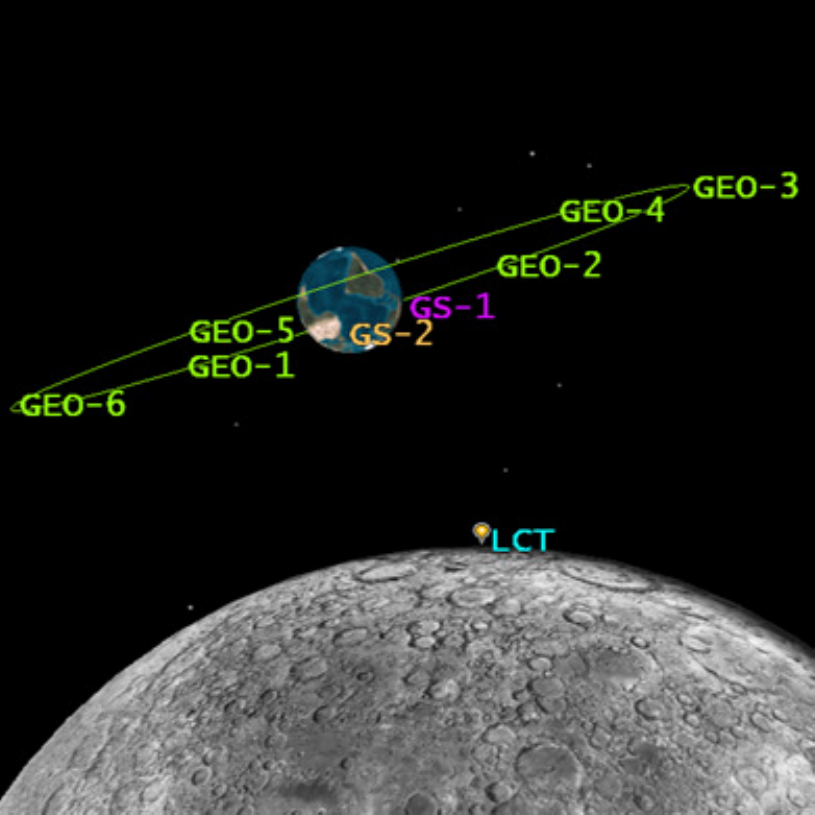}
\caption{Scenario~3}
\label{subfig_c}
\end{subfigure}\hspace*{\fill}
\begin{subfigure}{0.245\linewidth}
\centering
\includegraphics[width=\linewidth]{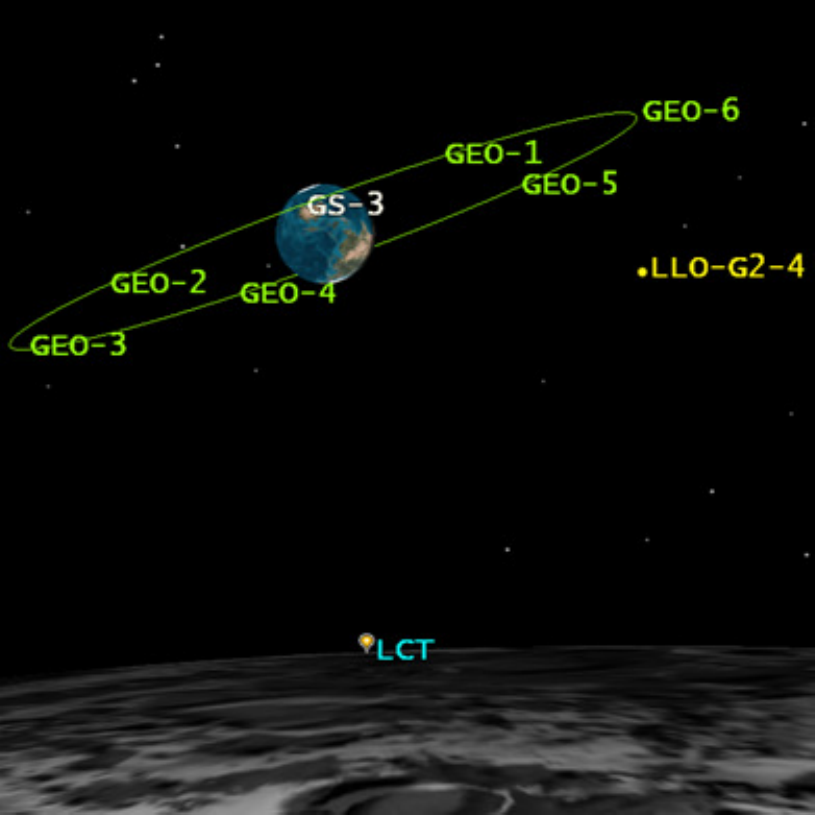}
\caption{Scenario~4}
\label{subfig_d}
\end{subfigure}
\caption{STK illustrations of communication scenarios correspond to four lunar network architectures. (a) Scenario~1: LCT transmits signals to GS on Earth. (b) Scenario~2: LCT sends signals to relay satellites in LLO, conveying them toward Earth. (c) Scenario~3: GEO satellites relay the signals received from LCT to the GS. (d) Scenario~4: Both LLO satellites and GEO satellites relay the signals received from LCT to the GS.}
\label{fig_stk}
\end{figure*}
\section{NUMERICAL RESULTS}\label{numeric_res}
In this section, we present the numerical results obtained from simulations based on the framework and strategies described in Sections \ref{sec_unified_framework}, \ref{inter-domain-space-networks} and \ref{dynamic-dt}. We first provide a detailed link analysis that is carried out on STK to ensure realistic orchestration of scenarios, as shown in Fig. \ref{fig_stk}. Then, we evaluate the reliability performance of each scenario and demonstrate the effectiveness of the digital twin algorithm to provide stable reliability.
\begin{table}[!b]
\centering
\caption{Link parameters of specified scenarios for lunar communication architectures \cite{ref1_t1,ref2_t1,ref4_t1,ref5_t1,ref7_t1}.}
\label{LinkBudgetParams}
\resizebox{\columnwidth}{!}{
\begin{tblr}{
cells = {c,m},
row{17} = {c},row{18} = {c},row{19} = {c},
column{5} = {c},column{6} = {c},column{7} = {c},column{8} = {c},
cell{1}{1} = {c=2}{c},cell{1}{3} = {c},cell{1}{4} = {c},
cell{2}{1} = {c=2}{c},cell{2}{3} = {c},cell{2}{4} = {c},
cell{3}{1} = {c=2}{c}{},cell{3}{4} = {c=6}{}, 
cell{4}{1} = {r=4}{c},cell{4}{2} = {c},cell{4}{3} = {c},cell{4}{4} = {c},
cell{5}{4} = {c}, cell{6}{4} = {c}, cell{7}{4} = {c},
cell{8}{1} = {r=7}{c},cell{8}{2} = {c},cell{8}{3} = {c},cell{8}{4} = {c},
cell{9}{4} = {c}, cell{10}{4} = {c}, cell{11}{4} = {c},
cell{12}{4} = {c}, cell{13}{4} = {c}, cell{14}{4} = {c},
cell{15}{1} = {c=2}{c},
cell{16}{1} = {c=2}{},cell{16}{4} = {c=6}{},
cell{17}{1} = {c=2}{c},
vlines, hline{1-4,8,15-18} = {-}{}, hline{5-7,9-15} = {2-9}{},
}
\textbf{Parameters}& &\textbf{Unit}
&{\textbf{Moon}\\\textbf{to}\\\textbf{Earth}}
&{\textbf{Moon}\\\textbf{to}\\\textbf{LLO}}
&{\textbf{LLO}\\\textbf{to}\\\textbf{Earth}}
&{\textbf{Moon}\\\textbf{to}\\\textbf{GEO}}
&{\textbf{GEO}\\\textbf{to}\\\textbf{Earth}}
&{\textbf{LLO}\\\textbf{to}\\\textbf{GEO}}\\
$f$           &              & GHz      &26.25   &27.25  &26.25   &27.25   &26.25   &27.25\\
$B$           &              & MHz      &50      &50     &50      &50      &50      &50    \\
\begin{sideways}\textbf{Transmitter}\end{sideways} 
              & $P_{T}$      & W        &2       &0.5    &2       &1.99    &0.01    &2   \\
              & $\eta_{o}$   &$-$       &0.85    &0.85   &0.90    &0.85    &0.93    &0.90\\
              & $D_{A}$      & m        &1.5     &1.5    &1.5     &1.5     &4.6     &1.5\\
              & $L_{T}$      & dB       &1.5     &1.5    &1       &1.5     &0.5     &1\\
\begin{sideways}\textbf{Receiver}\end{sideways}
              & $\eta_{o}$   &$-$       &0.95    &0.90   &0.95    &0.93    &0.95    &0.93\\
              & $D_{A}$      & m        &34      &1.5    &34      &4.6     &34      &4.6\\
              & $L_{R}$      & dB       &0.5     &1      &0.5     &0.5     &0.5     &0.5\\
              & $T_{R}$      & K        &50      &100    &50      &70      &50      &70\\
              & $T_{\mathrm{AP}}$     & K        &290     &250    &290     &270     &290     &270\\
              & $T_{\mathrm{TLP}}$    & K        &290     &250    &290     &270     &290     &270\\
              & $\eta_\mathrm{TL}$  &$-$       &0.95    &0.90   &0.95    &0.95    &0.95    &0.95\\
$T_\mathrm{ATMP}$ &         & K        &290     & $-$   &290     & $-$    &290     &$-$\\
$T_\mathrm{CMB}$  &         & K        &2.725   & 2.725 & 2.725  & 2.725  & 2.725  &2.725\\
Rician-$K$        &         & dB       &10      &20     &15      &13      &30      &40
\end{tblr}
}
\end{table}
\subsection{Simulation Set-up and Configuration}
\begin{figure*}[!t]
\centering
\includegraphics[width=\textwidth]{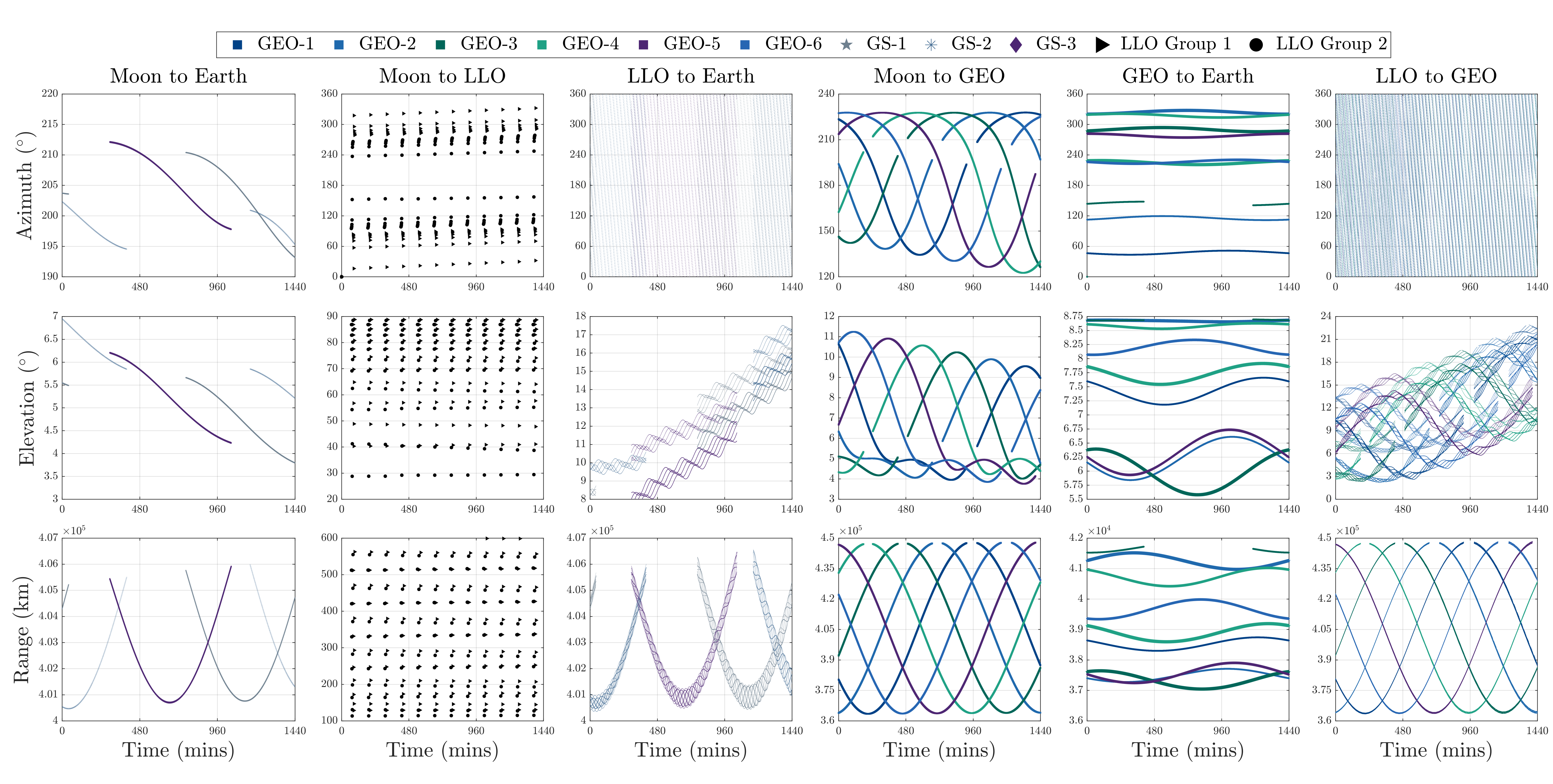}\vspace{2mm}
\caption{In-orbit measurements for the proposed scenario links for digital mission engineering, simulated with STK.}
\label{fig_AER}
\end{figure*}
The lunar communication architectures are simulated on STK through the four scenarios with a detailed link parameters in Table \ref{LinkBudgetParams}. The scenarios are evaluated minute-by-minute over a 24-hour period on October 1, 2024 with the following settings and configurations.
\begin{itemize}
\item The Mons Malapert ($\sim$$86^\circ\mathrm{S}, 0^\circ\mathrm{E}$)\footnote{The selenographic coordinate system is utilized to specify Mons Malapert's location on the lunar surface.}, a promising candidate for future missions because of its potential water ice presence and near-consistent illumination \cite{baris_hybrid}, hosts the lunar communication terminal~(LCT), as shown in Fig. \ref{fig_stk}.
\item Three DSN-NSN ground stations (GS) are located in Goldstone, California (GS-1); Canberra, Australia (GS-2); and Madrid, Spain (GS-3).
\item To maintain uninterrupted connectivity between the LLO network and the LCT, ten satellites are distributed equally across two orbital planes with inclinations of $84^\circ$ and $88^\circ$. 
\item In the beginning, LLO~group~1 satellites are set with true anomalies of $\left\{0^\circ,~72^\circ,~144^\circ,~216^\circ,~288^\circ \right\}$, whereas LLO~group~2 satellites are assigned true anomalies of $\left\{36^\circ,~108^\circ,~180^\circ,~252^\circ,~324^\circ \right\}$.
\item Six GEO relay satellites are deployed over the orbit, and the corresponding true anomaly values are set as $\left\{0^\circ,~60^\circ,~120^\circ,~180^\circ,~240^\circ,~300^\circ \right\}$, respectively.
\item Two-way communication is utilized by all relay satellites, which are equipped with antennas that can operate in both directions.
\item Since the Moon lacks an atmosphere like Earth's and the lunar regolith reflects light differently than on Earth, Moon's actual brightness temperature varies spatially and temporally. These illumination fluctuations may also impact receiver antennas pointing toward an LLO satellite, exposing them to the lunar surface in the background. To encompass this range of conditions in our analysis, the brightness temperature for antennas pointing toward the Moon or the LLO satellite was varied as: $T_B\in\left\{0,100,200,300,400 \right\}~\mathrm{K}$, while it is fixed at $150~\mathrm{K}$ for antennas pointing to GEO satellites.
\item The minimum SNR threshold is set to 10 dB, a value that supports high-fidelity communication for the robust modulation and coding schemes typically required for cislunar and deep space missions.
\item We established three targeted reliability levels: $\mathcal{L}=$~$\left\{\text{High}, \text{Moderate}, \text{Low}\right\}$ with corresponding maximum acceptable outage probability targets of $P_\mathrm{out}^l = \left\{10^{-5}, 10^{-4}, 10^{-3}\right\}$, respectively.
\end{itemize}
\subsection{In-Orbit Link Geometry and Propagation Dynamics}
We analyze how geometrical factors of transmission dynamics—azimuth, elevation, and range (AER)—vary for communication links corresponding to four scenarios throughout a day, as shown in Fig.~\ref{fig_AER}. These measurements are critical inputs for the unified link analysis framework, as they reveal significant variations in link conditions directly impacting communication visibility and signal propagation. Particularly, they highlight the challenges that necessitate inter-domain cooperation between space networks and their dynamic management.

We observe daily intervals of visibility and non-visibility for links in Fig.~\ref{fig_AER}, as well as acquire insight into LoS conditions through changes in azimuth and elevation. The elevation angles of \text{Moon-to-Earth} direct links are usually around five degrees, denoting that the Moon is barely above the horizon from the perspective of three~GS. Although there are still LoS, the fluctuating and low elevation means highly variable atmospheric path lengths for ground stations. When this is combined with ranges from roughly $370\,000$~km to over $400\,000$~km, the need for alternative relay solutions becomes clear. \text{Moon-to-LLO} links offer very short ranges, which are advantageous for minimizing path loss on the lunar access segment, but they have highly dynamic azimuth and elevation angles due to multiple LLO satellites. However, this benefit is countered by the rapid changes in azimuth and elevation, \textit{indicating a highly dynamic link and demanding precise antenna pointing and frequent handovers to maintain connectivity.} \text{LLO-to-Earth} links exhibit higher azimuth and elevation compared to the \text{Moon-to-Earth} link, but at larger distances. These longer and highly variable distances, coupled with dynamic angles, mean significantly fluctuating path loss.

\text{Moon-to-GEO} links show smooth and periodic variations in AER that imply a more stable propagation environment than LLO links. The high distance of GEO relays allows for broad coverage of the cislunar space, and their consistent visibility provides opportunities for handovers to maintain connectivity; however, the considerable range of approximately $360\,000$~km to $400\,000$~km poses a challenge. In contrast, \text{GEO-to-Earth} links provide relatively stable azimuth and elevation over long periods with a short range around $36\,000$~km. \textit{This means more predictable and lower free space path loss, which makes it more robust.} \text{LLO-to-GEO} links (scenario~4) also show periodic but dynamic changes in AER. The complexity of such multi-hop links is evident, as they combine the characteristics of several individual links, inheriting their distinct dynamic behaviors. \textit{The pervasive dynamism among these potential links underscores the requirement for intelligent, adaptive, and dynamic management of these architectures to ensure continuous communication and use resources effectively while maximizing their advantages.}
\begin{figure*}[!t]
\centering
\includegraphics[width=\linewidth]{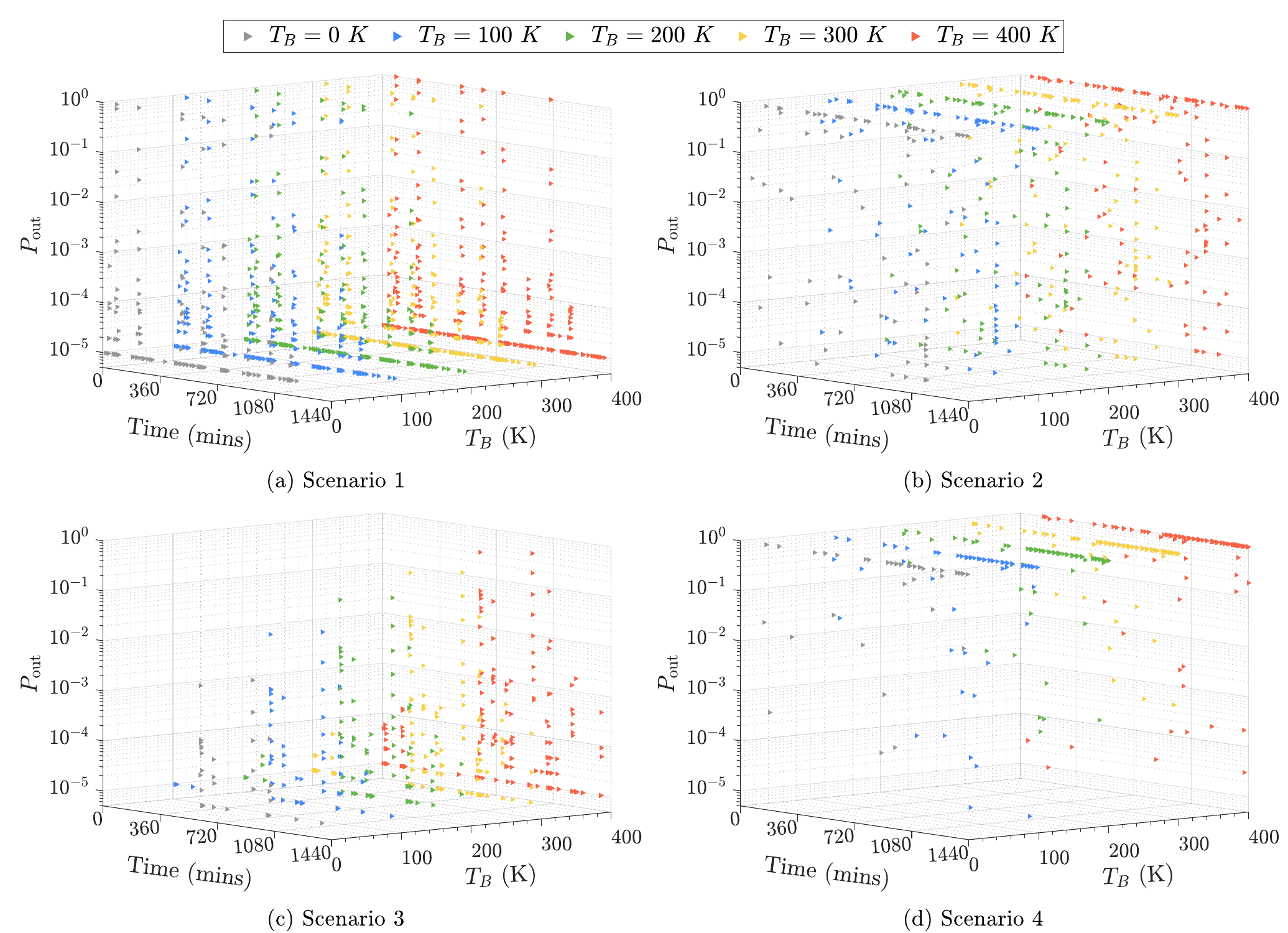}
\caption{The reliability of lunar architectures given their inherent dynamic link conditions via minute-by-minute variations in the outage probability for corresponding scenarios at $\gamma_\mathrm{th}=10$~dB.}
\label{outage_res}
\end{figure*}
\subsection{Architecture Reliability Assessment}
This section examines the reliability performance of four lunar communication architectures with corresponding scenarios, building upon the dynamic link conditions established in Fig.~\ref{fig_AER}. Fig.~\ref{outage_res} illustrates the outage probability for each scenario over time under different brightness temperatures and an SNR threshold of 10\,dB. Individual architecture analyses present their robustness for inter-domain space communication and implicitly lay the groundwork for the requirement of a dynamic decision-making mechanism among them. These analyses also allow us to address the trade-off between each scenario's architectural complexity and its resulting operational robustness.

While simple in its DTE network design, scenario~1 requires immense complexity on the ground to overcome LoS issues and the strong signal attenuation. This trade-off is compounded by its poor operational robustness, as shown in Fig.~\ref{outage_res}a. The outage probability for direct \text{Moon-to-Earth} communication exhibits significant variability and frequently reaches high values. This variability directly results from the visibility constraints observed in Fig.~\ref{fig_AER}, where the elevations are very low, despite the presence of LoS via three ground stations. The high outage probabilities, particularly during high ranges, demonstrate that conventional direct communication is not sufficiently reliable for continuous operations on future lunar missions. \textit{The impact of increasing illumination is visible; while it influences the noise floor, the primary driver of outage in this architecture is the fundamental geometric and visibility limitation.}

Scenario~2 adds architectural complexity with LLO relays to solve LoS issues in scenario~1, but it introduces new robustness challenges. In fact, the number of high outages is larger in Fig.~\ref{outage_res}b, especially mostly around the times of handover between LLO satellites and with higher $T_B$. This shows the dynamic nature of LLO links, as shown in Fig.~\ref{fig_AER}, where rapid changes in visibility and frequent handovers might lead to instant changes for the links or outages. \textit{LLO relays offer lower path loss and better visibility in the cislunar domain (due to closer proximity to the lunar surface); however, the overall reliability depends on the availability and stability of link hops in the architecture.} This means that maintaining a consistently low outage across both hops is challenging due to inherent orbital dynamics. 

Scenario~3, relying on bidirectional GEO relay, generally demonstrates a more stable and often lower outage probability compared to scenario~1 and scenario~2. The outage probabilities mostly fall at $10^{-3}$ and below, indicating significantly higher reliability. \textit{The smooth and periodic link dynamics of GEO satellites, as observed in~Fig.~\ref{fig_AER}, contribute to this improved stability.} While our analysis uses complex DSN ground stations for all scenarios to ensure a fair comparison—making scenario~3 initially appear more complex—a DSN ground station is not essential for the stable GEO-to-Earth link. This highlights another potential feasibility of the architecture in practice: using simpler NSN ground stations to offload immense hardware complexity from the ground, thereby creating a more scalable solution. Although the outage probability increases with higher $T_B$ values, the overall reliability remains lower than in other scenarios. Both results, in Fig.~\ref{fig_AER} and Fig.~\ref{outage_res}c, justify the advantages of the proposed NSN architecture for maintaining consistent and robust connectivity over large~distances.

As the most complex architecture, the multi-hop scenario~4 generally exhibits the lowest outage probabilities among all scenarios across various $T_B$ values, except for the highest outages caused during the handover of LLO relay satellites, as shown in Fig.~\ref{outage_res}d. This justifies that if this problem is overcome (such as by using more LLO satellites), the architecture will provide superior reliability by leveraging the benefits of both LLO (for short ranges and better visibility) and GEO for robust backhaul to Earth (stable and predictable link dynamics). On the other hand, boosting the multi-hop design with more satellites inherently means more redundancy, complexity, and resource usage. Therefore, \textit{the challenging trade-off between performance and cost constraints leads to scenario~3 being more amenable, while the higher performance demands make scenario~4~superior.}

Across all scenarios, an increase in the brightness temperature typically correlates with an increase in outage probability. This is because $T_B$ contributes to the external antenna noise temperature, which in turn increases the operational equivalent noise temperature and thus the total noise power at the receiver. \textit{The sensitivity of outage probability to lunar surface brightness emphasizes the importance of accurately modeling the noise environment.} 

While our analysis assumes a uniform surface brightness temperature for the 24-hour period since the average temperature at a location like Mons Malapert does not fluctuate dramatically within this short time frame, the broader context of the 29.5-day lunar cycle would impose slow variations on $T_B$. Furthermore, the real-world lunar thermal environment is complicated by spatial and topographical variations that create a complex thermal map. Although the LCT is not mobile, its location at Mons Malapert near the South Pole is an area known for long shadows and sharp temperature gradients. An antenna tracking a satellite could therefore sweep its view from a sunlit patch to a permanently shadowed crater, causing rapid and unpredictable fluctuations in the external antenna noise temperature. These thermal hot spots would impact the outage probability momentarily due to noise rather than link geometry alone. These real-world complexities create a critical gap that static communication architectures cannot fill.

This is precisely the motivation for the development of the inter-domain space digital twin. The results show that no single architecture is always ideal, and a dynamically adaptive solution is necessary to overcome the inherent trade-offs between them. The digital twin is designed not just to select the best path based on predictable orbital dynamics but to proactively mitigate these time- and location-dependent noise variations. By integrating real-time telemetry with predictive models, the digital twin can switch to a more robust communication path before noise compromises the primary link, ensuring the unwavering reliability that future lunar missions demand.
\subsection{Validation of Dynamic Orchestration}
We observe the dynamic and complicated nature of lunar communication with significant variability in link geometries and challenges like limited line-of-sight, rapidly changing LLO relay links, and path losses via Fig.~\ref{fig_AER}. Fig.~\ref{outage_res} further reveals no single architecture consistently provides optimal reliability. This comprehensive analysis establishes that a dynamic orchestration is indispensable for continuous, robust, and resource-efficient lunar communication. Therefore, we introduce an inter-domain space digital twin that continuously models the entire communication, dynamically optimizing strategies by balancing unwavering reliability with efficient transmit power utilization.
\begin{table}[!t]
\vspace{-2mm}
\centering
\caption{Daily total outage duration in minutes and average power consumption of lunar architectures with and without dynamic management by digital twin.}
\label{table2}
\resizebox{\columnwidth}{!}{\huge
\begin{tblr}{
  cells = {c,m},
  cell{2}{1} = {c=6}{},
  cell{4}{2} = {c=5}{},
  cell{5}{1} = {c=6}{},
  cell{7}{2} = {c=5}{},
  cell{8}{1} = {c=6}{},
  cell{10}{2} = {c=5}{},
  cell{11}{1} = {c=6}{},
  cell{13}{2} = {c=5}{},
  cell{14}{1} = {c=6}{},
  hline{1,17} = {-}{0.08em},
  hline{2,5,8,11,14} = {-}{0.05em},
}
{Metrics}                               & $T_B = 0$  &$T_B = 100$& $T_B = 200$ & $T_B = 300$  & $T_B = 400$ \\
{Scenario~1} (Direct to Earth)   &            &   &   &   &  \\
{Total Outage}                          & 10.2284    & 12.1227   & 13.6884     & 15.0746      & 16.3272\\
Avg. $P_{T,\mathrm{total}}$                 & 2          &   &   &   &  \\
{Scenario~2} (via an LLO relay)  &            &   &   &   &  \\
{Total Outage}                          & 37.8367    & 45.2522   & 51.6477     & 56.6411      & 61.0316 \\
Avg. $P_{T,\mathrm{total}}$                 & 2.5        &   &   &   &  \\
{Scenario~3} (via a GEO relay)   &            &   &   &   &  \\
{Total Outage}                          & 0.0102     & 0.0725    & 0.2720      & 0.7117       & 1.4501\\
Avg. $P_{T,\mathrm{total}}$                 & 2          &   &   &   &  \\
{Scenario~4} (via an LLO and a GEO relay)&    &   &   &   &  \\
{Total Outage}                          & 30.9685    & 49.5117   & 73.6876     & 94.4604      & 119.6084\\
Avg.$P_{T,\mathrm{total}}$                 & 2.51       &   &   &   &  \\
{All Scenarios with Digital Twin Management}  &   &   &   &   &       \\
{Total Outage}                          &0.00013     &0.00026    &0.00042      &0.00073       &0.0013\\
Avg. $P_{T,\mathrm{total}}$                 & 2.0007     &2.0014     &2.0032       &2.0046        &2.0060                                         
\end{tblr}}
\end{table}
\begin{figure*}[!t]
\centering
\begin{subfigure}{0.9985\textwidth}
\centering
\includegraphics[width=\linewidth]{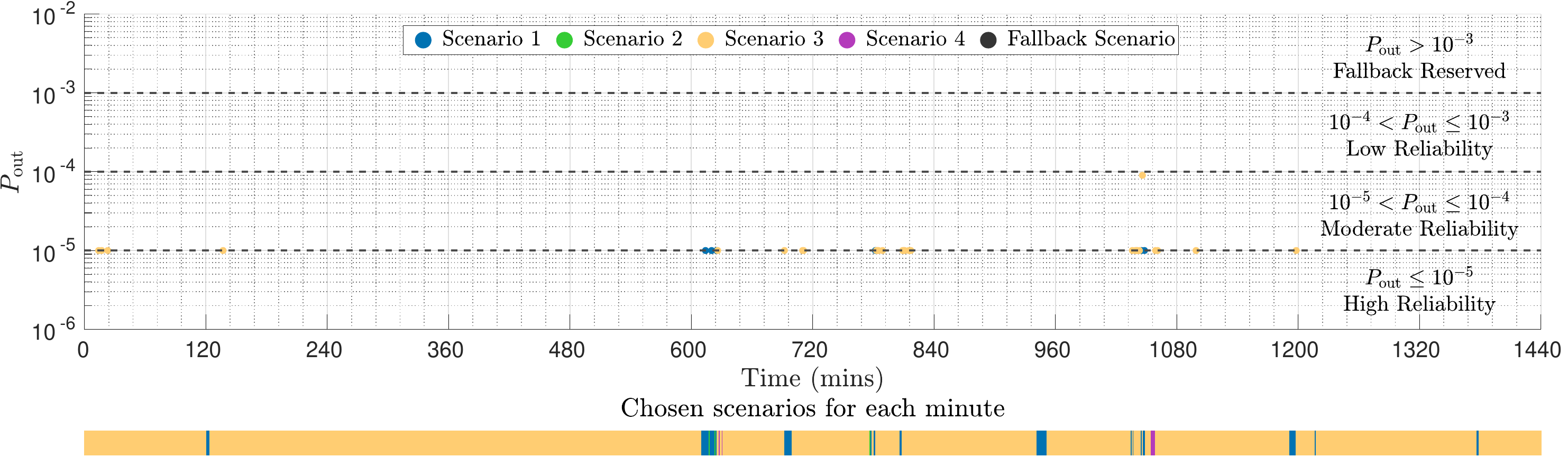}
\caption{$T_B = 200~\mathrm{K}$}
\label{twin200}
\end{subfigure}\\
\begin{subfigure}{0.9985\textwidth}
\centering
\includegraphics[width=\linewidth]{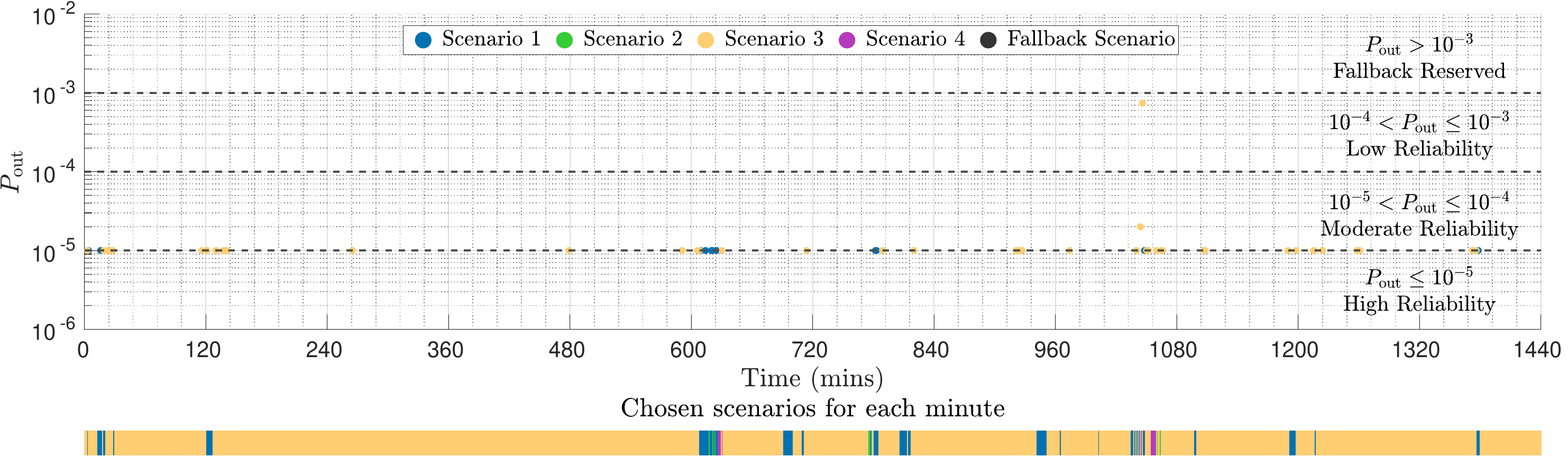}
\caption{$T_B = 400~\mathrm{K}$}
\label{twin400}
\end{subfigure}
\caption{The dynamic orchestration of scenarios by digital twin to sustain robust communication. Scenario~3 is set as the reserved fallback scenario for the digital twin, given its performance among the individual architectures.}
\label{twin}
\end{figure*}

Table \ref{table2} presents a quantitative summary of the daily total outage duration and average total transmit power ($P_{T,\mathrm{total}}$) over the 24-hour period. The results clearly illustrate the significant disparity in operational robustness among the individual architectures and reveal the profound improvements achieved through the dynamic orchestration of the proposed digital twin. Compared to its counterparts, the third scenario (based on GEO relaying) demonstrates the highest reliability, with the lowest outage duration for an average $P_{T,\mathrm{total}}$ of 2~W. Despite its superior performance, uninterrupted connectivity is essential for cislunar missions, where even an outage of a few seconds can have critical consequences. Here, the digital twin's role becomes mandatory: under challenging noise conditions ($T_B=400~\mathrm{K}$), it reduces the total daily outage from scenario~3's 87 seconds to just 0.078 seconds, a significant gain in reduction of over 99.9\%, all for a negligible 0.3\% increase in average total power. In other words, it increases the daily link availability from 99.90\% for the best static scenario to 99.9999\%. This result underscores that dynamic architectural management is essential for inter-domain space networks, allowing the system to adapt to changing conditions and always select the most reliable and power-efficient communication path. The subsequent analysis addresses precisely how the digital twin achieves the dramatic performance improvements in Table \ref{table2}.

Fig.~\ref{twin} compellingly illustrates this minute-by-minute orchestration, showing how the digital twin sustains a highly reliable link throughout the day under different thermal noise conditions. For a brightness temperature of 200~K, as shown in Fig.~\ref{twin200}, the digital twin maintains the communication link predominantly within the high-reliability target ($P_\mathrm{out} \le 10^{-5}$). The lower bar reveals the system's underlying strategy: while it generally relies on the robust and relatively simple scenarios~1~and~3, its sophistication is evident in its tactical use of more complex multi-hop architectures. For instance, in the window between minutes 1032 and 1080, a period where scenarios~1~and~3 face a higher risk of outage, the digital twin switches to scenario~4 to ensure uninterrupted connectivity. This decision illustrates the calculated trade-off at the core of the engine's logic: it embraces higher architectural complexity only when it is essential for maintaining robustness, while simultaneously selecting the most power-efficient path that meets the reliability target. This adaptable balancing of reliability against complexity and power consumption is crucial for enabling stable real-time command and control in the resource-scarce cislunar domain. Even when conditions cause brief dips into moderate reliability, the system quickly recovers, showcasing its ability to leverage the strengths of each architecture in response to the dynamic environment illustrated earlier in Fig.~\ref{fig_AER}.

To further validate the digital twin's resilience, Fig.~\ref{twin400} shows its performance under the more challenging condition of an increased brightness temperature of 400~K, where the engine continues to strive for high reliability. While largely operating within the high-reliability level, more frequent and slightly longer periods of outage probability fall into moderate or low-reliability levels. This behavior is a direct result of the elevated noise floor accentuating the inherent architectural trade-offs discussed previously. For instance, the immense path loss associated with the otherwise stable Moon-to-GEO link (scenario~3) becomes more critical for the digital twin's decision. Simultaneously, the system becomes more sensitive to the highly dynamic nature of LLO links, where outages are most common during the frequent satellite handovers required for continuous connectivity (a key weakness of scenarios~2~and~4). However, the digital twin adapts to more challenging noise environments and ensures robustness by dynamically selecting scenarios to mitigate performance degradation. Crucially, even as the system is pushed into moderate or low reliability more often, it successfully avoids the fallback reserved level entirely. This dynamic adaptability allows the system to effectively manage the trade-offs exacerbated by the high noise floor. It ensures unwavering reliability—a paramount factor for mission success and safety—by intelligently selecting the best available path. This is essential, as it provides robust connectivity even when scarce power or geometric constraints make achieving the absolute highest reliability impractical.

Overall, the inter-domain space digital twin provides a groundbreaking solution for advancing lunar communication. By continuously assessing link conditions, prioritizing unwavering reliability, and optimizing power consumption through dynamic scenario selection, it effectively overcomes the inherent limitations and complexities of individual architectures. This adaptive capability is fundamental for the success, safety, and scalability required for future lunar endeavors.
\section{CONCLUSIONS}
Establishing and maintaining reliable communication links is of paramount importance for lunar missions, as every phase of a mission, from telemetry and command to astronaut safety, depends on robust and uninterrupted connectivity. Failures or significant degradation in communication quality can jeopardize mission objectives and even endanger assets or human lives. This critical need for unwavering reliability motivates a move beyond conventional DTE communication, which is plagued by LoS limitations, economic non-scalability, and scheduling bottlenecks. This paper addresses these challenges by proposing a foundational shift towards inter-domain space network cooperation, integrating CSNs with NSNs. 

To validate and compare these advanced architectures, we developed a unified link analysis framework that provides a high-fidelity assessment by incorporating unique environmental factors. Second, building on this framework, we assessed their reliability via outage probability to quantify the operational robustness of architectures under dynamic conditions. Through these, our work shows the advantages of NSN integrated architectures but also underscores the emerging requirement to build a resilient, dynamic, and high-performance communication backbone. Therefore, we proposed the inter-domain space digital twin, a dynamic decision-making engine that autonomously analyzes the networks in real-time to dynamically select the most reliable and power-efficient communication path. Final results show that the proposed approach ensures optimal performance and resource utilization for the next generation of sustained lunar exploration.
\bibliographystyle{IEEEtran}
\bibliography{IEEEtaes}
\begin{IEEEbiography}[{\includegraphics[width=1in,height=1.5in,clip,keepaspectratio]{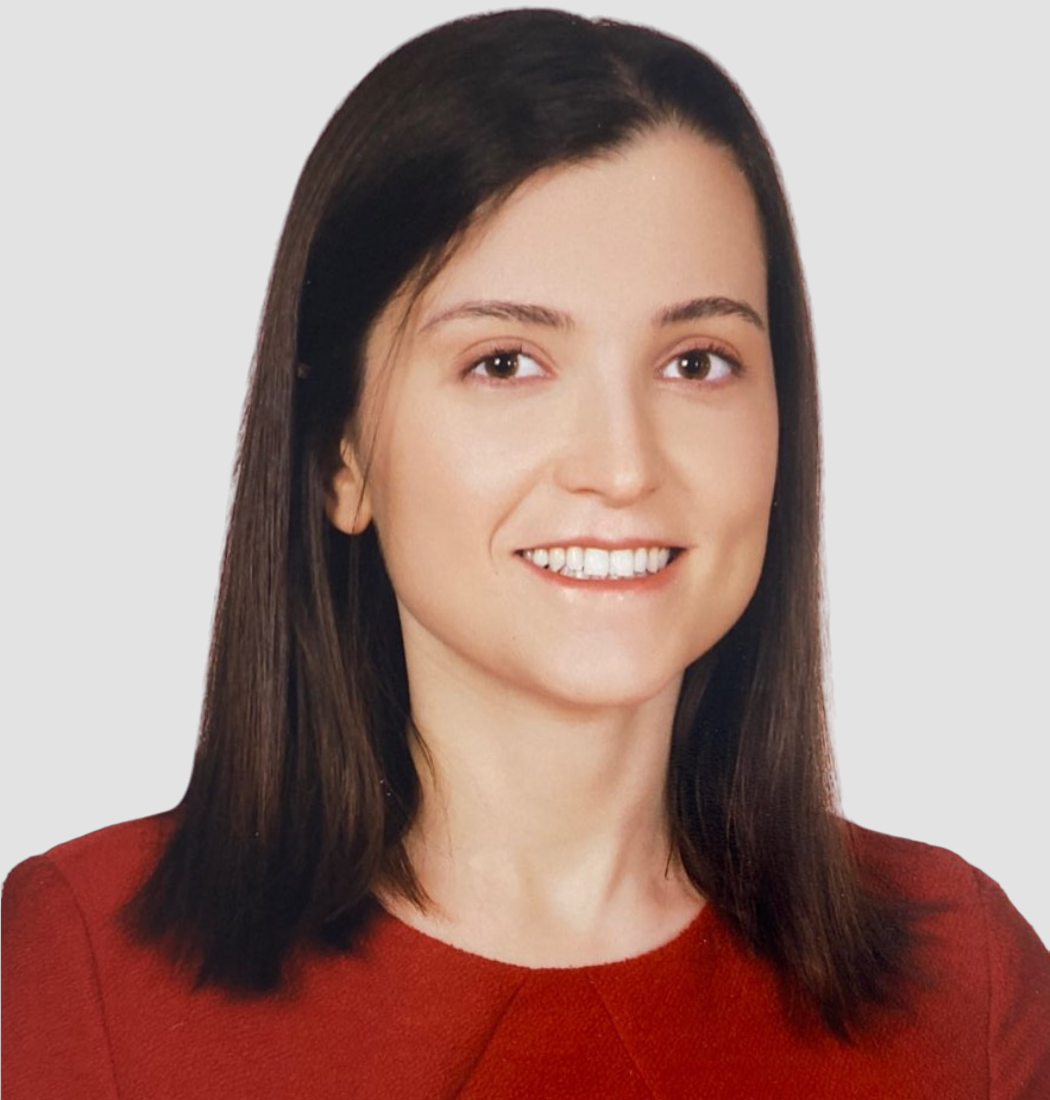}}]{Selen Geçgel Çetin}{\space}(Member, IEEE)~received the B.S. degree from Yildiz Technical University in 2016, and the M.S. and Ph.D. degrees from Istanbul Technical University (ITU) in 2019 and 2025, respectively. She is a member of the Wireless Communication Research Laboratory at ITU and the Poly-Grames Research Center at Polytechnique Montréal. Her research focuses on the intersection of machine learning and wireless communications, with a specific emphasis on intelligent designing and adaptive managing for secure and reliable space communication.
\end{IEEEbiography}
\begin{IEEEbiography}[{\includegraphics[width=1in,height=1.5in,clip,keepaspectratio]{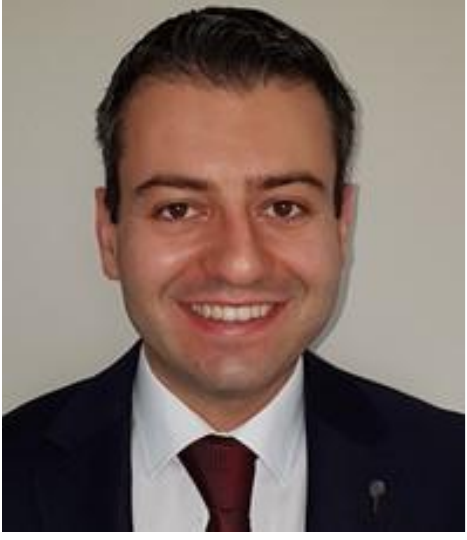}}]{Barış Dönmez}{\space}{\space}(Graduate Student Member, IEEE)~received the B.Sc. and M.Sc. degrees (with high honors) in electrical and electronics engineering, in 2009 and 2022, respectively, from FMV Işık University, Istanbul, Turkiye. He is currently pursuing his Ph.D. degree in electrical engineering at Polytechnique Montréal, Montréal, QC, Canada. His research interests include communication and energy harvesting systems in space networks.
\end{IEEEbiography}
\begin{IEEEbiography}
[{\includegraphics[width=1in,height=1.5in,clip,keepaspectratio]{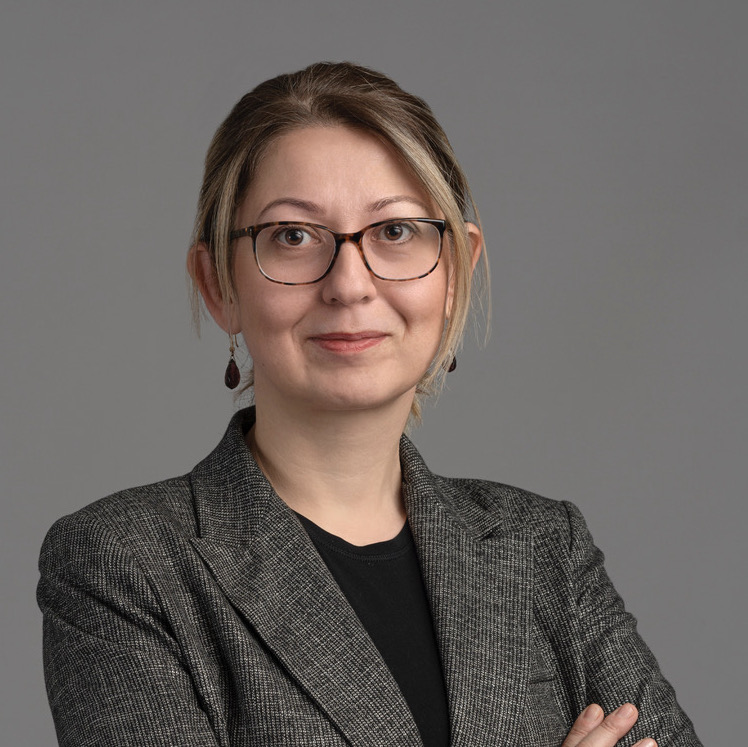}}] {Güneş Karabulut Kurt}{\space} [M’00, SM’14] is a Canada Research Chair (Tier 1) in New Frontiers in Space Communications and Professor at Polytechnique Montréal, Montréal, QC, Canada. She is also an adjunct research professor at Carleton University, ON, Canada. Güneş is a Marie Curie Fellow and has received the Turkish Academy of Sciences Outstanding Young Scientist (TÜBA-GEBIP) Award in 2019. She received her Ph.D. degree in Electrical Engineering from the University of Ottawa, ON, Canada.
\end{IEEEbiography}
\end{document}